\documentclass[epj,referee]{svjour}
\usepackage{graphicx}
\begin{document}
\title{Electronic structure and magnetic properties of $R$Mn$X$ 
($R$ = Mg, Ca, Sr, Ba, Y; $X$= Si, Ge) studied by KKR method}
\author{V. Klosek\inst{1} \and J. Tobo\l{}a\inst{1,2} \and 
A. Verni\`{e}re\inst{1} \and S. Kaprzyk\inst{2} 
\and B. Malaman\inst{1}
%
\thanks{\emph{Corresponding author:} 
Bernard.Malaman@lcsm.uhp-nancy.fr}%
}                     
%
%
\institute{Laboratoire de Chimie du Solide Min\'{e}ral, Universit\'{e} 
Henri Poincar\'{e} - Nancy I, Associ\'{e} au CNRS (UMR 7555), 
BP 239, F-54506 Vandoeuvre-l\`{e}s-Nancy Cedex, France 
\and Faculty of Physics and Nuclear Techniques,
AGH University of Science and Technology, Al. Mickiewicza 30, 
30-059 Krak\'ow, Poland
}
\date{Received: date / Revised version: date}
%
\abstract{Electronic structure calculations, using the charge and 
spin self-consistent 
Korringa- Kohn-Rostoker (KKR) method, have been performed for several 
$R$Mn$X$ compounds ($R$ = Mg, Ca, Sr, Ba, Y; $X$ = Si, Ge) of the CeFeSi-type 
structure. 
The origin of their magnetic properties has been investigated emphasizing 
the role of the Mn sublattice. The significant influence of the Mn-Mn and 
Mn-$X$ interatomic distances on the Mn magnetic moment value is delineated 
from our computations, supporting many neutron diffraction data. 
We show that the marked change of $\mu_{Mn}$ with the Mn-Mn and Mn-$X$ 
distances resulted from a redistribution between spin-up and 
spin-down $d$-Mn DOS rather than from different fillings of the Mn 
3$d$-shell. The obtained KKR results are discussed considering 
the Stoner-like and covalent magnetism effects. 
From comparison of electronic structure of $R$Mn$X$ in different 
magnetic states we conclude that the antiferromagnetic coupling in 
the Mn (001) plane considerably increases the Mn magnetic moment with 
respect to the ferromagnetic arrangement. 
Bearing in mind that the neutron diffraction data reported for the $R$Mn$X$ 
compounds are rather scattered, the KKR computations of $\mu_{Mn}$ are in 
fair agreement with the experimental values. 
Comparing density of states near $E_{F}$ obtained in different 
magnetic orderings, one can notice that the entitled $R$Mn$X$ systems 
seem to 'adapt' their magnetic structures to minimize the DOS in the 
vicinity of the Fermi level. Noteworthy, the SrMnGe antiferromagnet 
exhibits a pseudo-gap behaviour at $E_{F}$, suggesting anomalous 
electron transport properties.
In addition, the F-AF transition occurring in the disordered 
La$_{1-x}$Y$_{x}$MnSi alloy for the $0.8<x<1$ range is well supported 
by the DOS features of La$_{0.2}$Y$_{0.8}$MnSi. 
This latter result sheds light on the magnetic structure of the YMnSi 
compound. 
In contrast to the investigated $R$Mn$X$ compounds, YFeSi was found to 
be non-magnetic, which is in excellent agreement 
with the experimental data.
} 
\maketitle
\section{Introduction}

Among the equiatomic $RTX$ compounds ($R$ = rare or alkaline earth; 
$T$ = transition element; $X$ = metalloid), manganese is the only $d$ 
element that exhibits a magnetic moment \cite{szytula1,szytula2}. 
Thus $R$Mn$X$ representatives allow for investigations of 3$d$ itinerant magnetism in the $RTX$ phases.

The structural and magnetic properties of the CeFeSi-type $R$Mn$X$ 
($R$ = La - Tb (or Sm), Y, Mg - Ba) silicides and germanides have been 
extensively studied during the last decade, especially by means of $X$-ray 
diffraction, magnetic measurements, and neutron powder diffraction 
\cite{szytula1,szytula2,kido3,welter4,welter7,welter8,ijjaali9,dascou10,welter11,klosek12}. 
The CeFeSi-type structure (Table~\ref{table1}, Fig.~\ref{fig1}) is 
closely related to the well-known ThCr$_{2}$Si$_{2}$-type structure. 
Both CeFeSi- (P4/$nmm$) and ThCr$_{2}$Si$_{2}$- (I4/$mmm$) types are 
characterised by the occurrence of 'BaAl4' slabs which 
consist in a stacking along the c axis of $R$, $X$, and $T_{2}$ (001) 
square planes with the sequence $R-X-T_{2}-X-R$ (Fig.~\ref{fig1}). 
In the CeFeSi structure, the 'BaAl$_{4}$' slabs are connected via 
$R-R$ contacts constituting a 'W' slab. 
The $T$ atom ($CN 12$) has a close coordination of 
four $X$ atoms building a tetrahedron, and a second coordination shell 
constituted of four $R$ atoms building a second tetrahedron. 
Four $T$ atoms, building a square plane, complete the 
neighbourhood of $T$.

Neutron diffraction studies have shown that, in both CeFeSi-type $R$Mn$X$ 
and ThCr$_{2}$Si$_{2}$-type $R$Mn$_{2}$X$_{2}$ compounds, the magnetic 
behaviour of the Mn sublattice strongly depends on the Mn-Mn (d$_{Mn-Mn}$) 
and Mn-$X$ ($d_{Mn-X}$) interatomic distances. 
From many experimental studies one can notice that, within the (001) 
planes, the Mn-Mn coupling is ferromagnetic (F) for $d_{Mn-Mn} < 2.84$ \AA~
and antiferromagnetic (AF) for $d_{Mn-Mn} > 2.89$ \AA~whereas 
complex magnetic arrangement (coexistence of F and AF components) 
occurs for intermediate distances. Nevertheless, the case of MgMnGe 
($d_{Mn-Mn} = 2.79$ \AA~and AF Mn (001) planes) partly questions this metric 
criterion. On the other hand, the evolution of the $\mu_{Mn}$ moment in the 
$R$Mn$X$ compounds is well interpreted when plotting $\mu_{Mn}$ versus the 
contraction rate ({\it with respect to the sum of the Mn and X ionic radii}) 
of $d_{Mn-X}$ distance [see Fig. 6 of Ref.~\cite{welter7}]. 
The largest values of the Mn magnetic moment are observed in the 
compounds characterised by the largest Mn-$X$ and Mn-Mn distances 
\cite{welter4,welter8,ijjaali9,dascou10}.

The aim of this paper is to investigate theoretically the magnetic 
properties of the Mn sublattice in the CeFeSi-type $R$Mn$X$ silicides and 
germanides, from electronic structure calculations. We mainly focus 
on the dependency of the Mn magnetic moment value versus the Mn-Mn 
and Mn-$X$ interatomic distances as well as on the relationships between 
the electronic states in the vicinity of the Fermi level (E$_{F}$) 
and the magnetic structure type.

Furthermore, the influence of the nature of the transition metal on 
magnetic properties onset has been analysed through electronic structure 
calculations of the Y$T$Si ($T$= Cr, Mn, Fe) compounds.

\section{Computational details}

The electronic structure and ground state properties of the $R$Mn$X$ 
silicides and germanides have been calculated using 
the Korringa-Kohn-Rostoker (KKR) method \cite{bansil13,bansil14,kaprzyk15}. 
Both core states and valence states are calculated self-consistently. 
The crystal potential of muffin-tin (MT) form is constructed within 
the local spin density approximation (LSDA) with the 
Barth-Hedin expression for the exchange-correlation potential \cite{barth16}. 
The self-consistency cycles is repeated for each compound until the 
difference between the input and output crystal potentials is of about 
1 mRy.

We have carried out the KKR calculations for the MgMnGe, CaMnGe, SrMnGe, 
BaMnGe, CaMnSi and YMnSi compounds in (i) ferromagnetic structure (F) 
(ii) three types of antiferromagnetic structures namely antiferromagnetic 
Mn (001) planes coupled either antiferromagnetically (AFI) or 
ferromagnetically (AFII) (see below), ferromagnetic Mn (001) planes coupled 
antiferromagnetically (AFIII). The non-spin polarised DOS have been also 
analysed to compare the Mn-DOS at $E_{F}$ for the various Mn-Mn distances 
observed in the studied $R$Mn$X$ compounds. 
In the case of BaMnGe relativistic effects have been incorporated 
to compute core electronic states.

In the KKR calculations, the low temperature experimental values of 
the lattice parameters have been taken from the neutron diffraction 
data \cite{welter4,welter7,dascou10}, whereas the atomic 
coordinates have been taken 
from the single crystal X-ray diffraction data \cite{pearson} 
(Table~\ref{table2}). 
Noteworthy, for the YMnSi compound, the room temperature lattice 
parameters have been applied. Concerning the atomic coordinates, 
the values previously suggested for this compound (in fact those 
of the CeFeSi compound \cite{bodak17}) give a Mn-Si distance 
surprisingly short (2.35 \AA) comparing to those observed for 
other $R$MnSi compounds \cite{welter8}. So, in the following, we have used 
atomic positions extrapolated from Ref. \cite{welter8}, namely $z_{Y}=0.67$ 
and $z_{Si}=0.22$. In order to support this assumption, electronic 
structure of the La$_{0.2}$Y$_{0.8}$MnSi alloy (d$_{Mn-Si} = 2.50$ \AA) 
\cite{ijjaali9} was also studied by the KKR method incorporating the 
coherent potential approximation (CPA) to treat disorder.

The radii of the non-overlapping muffin-tin spheres have been chosen to 
obtain an optimal filling of the Wigner-Seitz cell (about 65-70\%). 
One has to notice that moderate variations of the computational geometry 
did not affect our results. For the final crystal potentials, the total, 
site- and $l$-decomposed (with $l_{max} = 2$) densities of states (DOS) 
have been computed on a 601-energy-point mesh with a tetrahedral k-space 
integration technique using 375 small tetrahedra and 126 {\bf k} points 
in the irreducible part of the Brillouin zone \cite{kaprzyk18}. 
In the case of SrMnGe, electronic dispersion curves E({\bf k}) have 
been computed along high symmetry directions in the Brillouin zone. 

The KKR-CPA cycles in La$_{0.2}$Y$_{0.8}$MnSi have been carried out in 
the complex energy plane using an elliptic contour and the Fermi level 
has been determined precisely via the generalized Lloyd formula 
\cite{kaprzyk19}.

In the presented figures the Fermi level ($E_{F}$) is taken as zero 
of the energy scale.

\section{Results and discussion}

\subsection{Non spin-polarised DOS}

The DOS curves of germanides and silicides obtained from non-spin-polarised 
calculations are displayed in Fig.~\ref{fig2}. The lowest energy bands 
are mostly attributed to the metalloid $s$-states and are separated from 
the upper-lying states by a gap of about 0.2-0.3 Ry. These $s$-states 
are weakly hybridised with the other electronic states and all $l$-like 
DOS (seen on the $R$ and Mn sites). They should be considered as the 
'tails' of the $X$ $s$-states and their role in bond formation can 
be neglected \cite{gelatt21,matar22,bisi23}. Moreover, we can note that the 
energy localization of the $s$-states increases when the size of the 
$R$ element increases from Mg to Ba (Fig.~\ref{fig2}). The expansion of the 
unit cell volume implies the increase of the $X-X$ interatomic distances 
and then the decrease of the overlaps of the $X$ $s$-orbitals. Above the gap, 
the valence states are mostly of Mn 3$d$ character. The Mn 3$d$-states 
are strongly hybridised with the $X$ $p$-states in the energy range of 
$-0.3 < E <-0.1$ Ry, yielding bonding states for Mn-$X$ bonds. 
The corresponding antibonding states form a peak located c.a. 0.1 Ry above 
E$_{F}$. Note that the $X$ $p$-orbitals contribute significantly to 
the bonding states while the antibonding states are essentially of 
Mn 3$d$ character. This may originate from the energy barycentre 
of the $X$ $p$-states lying lower than the one corresponding to the Mn 3$d$ 
states. The electronic states occurring in the vicinity of E$_{F}$ mostly 
come from direct overlaps between the Mn 3$d$ orbitals 
\cite{matar22,bisi23,klosek20}, corresponding to the (001) Mn planes 
(Fig.~\ref{fig1}). These states are located within a relatively narrow energy range, 
and there is no clear separation between bonding and antibonding states. 
The Mn-Mn bonds thus have a quite important metallic 
character and, anyway, they are less covalent than the Ti-Ti bonds 
calculated in the isotypic $R$Ti$X$ compounds (and even the 
Fe-Fe bonds, in the $R$FeSi compounds, see below) \cite{klosek20}.

The electronic states projected on $R$ sites give a little contribution 
to the density of states at the Fermi level. These states are dominated 
(except for MgMnGe) by $d$-states at higher energy range while there is a 
small admixture of $s$- and $p$-states at the lower energies. The hybridisation 
between the $R$ $s$-, $p$- and $d$-states and the $X$ $p$-states is 
rather important and there should be some bonding states for $R-X$ bonds. 
We can also note the hybridisation of $R$ electronic states with the Mn 3$d$ 
ones. This is particularly well seen in the case of the trivalent $R$ 
element (YMnSi), where the shape of the 4$d$ Y and 3$d$ Mn DOS is quite 
similar in the energy range $-0.3 < E < + 0.1$ Ry thus evidencing bonding 
states for Y-Mn bonds \cite{weijs24,johansson25}. Conversely, the $R$-Mn bonds must be 
weaker in the $R$Mn$X$ compounds containing the divalent $R$ elements, 
since the bonding states related to the $R$-Mn bonds are less filled.

In the case of YMnSi, the KKR results can be compared to the previous 
non-spin polarised LMTO calculations of Kulatov et al. \cite{kulatov26} 
which are, to our knowledge, the only band theory results 
reported for the $R$Mn$X$ 
compounds. Despite of the overall agreement concerning the shape of 
non-polarised DOS, the present KKR calculations give a larger 
value of $n(E_{F})$. However, we are not aware of the 
crystallographic data used in the previous LMTO study. 

Conclusively, the non-polarised KKR calculations of $R$Mn$X$ 
compounds show that the Fermi level is always located in the vicinity of 
large DOS peak originating essentially from 3$d$-states on Mn. 
Moreover, the comparison of the computed DOS at $E_{F}$ versus interatomic 
distances in $R$Mn$X$ compounds (Table~\ref{table3}) reveals that $n(E_{F})$ 
increases regularly with increasing d$_{Mn-Mn}$ (i.e. with the $R$ size) 
but there is no apparent correlation between $n(E_{F})$ and d$_{Mn-X}$. 
The electronic structure modifications with $d_{Mn-Mn}$ can be roughly 
understood by smaller overlap of 3$d$ orbitals, resulting in narrower 
bands and then enhanced density of states in the vicinity of the Fermi 
level. 

In all investigated $R$Mn$X$ compounds, the computed $n(E_{F})$ values for 
the Mn site are sufficiently large to satisfy the well-known Stoner 
limit ($I  n(E_{F}) >1$) for appearance of magnetism (Table~\ref{table3}). 
This may tentatively support the magnetic properties of the $R$Mn$X$ 
compounds ($R$= Y, Mg-Ba; $X$= Si, Ge) and led us to perform spin-polarised 
electronic structure calculations accounting for experimentally observed 
magnetic structures \cite{kido3,welter4,welter7,dascou10,klosek12}.

\subsubsection{Non-magnetic properties of $R$FeSi}

The origin of disappearance of magnetic properties in the $R$Fe$X$ compounds, 
unlike the aforementioned $R$Mn$X$ ones, can be tentatively illustrated 
when comparing e.g. the electronic structures of isotypic YFeSi, YMnSi 
and hypothetical YCrSi compounds (Fig.~\ref{fig3}). 

Using the lattice parameters of Ref.~\cite{pearson} 
($a = 3.89$ \AA, $c= 6.80$ \AA) and the atomic positions extrapolated 
from Ref.~\cite{welter8} ($z_{Y} =0.672$ 
and $z_{Si} =0.185$), the KKR computations of YFeSi show that $E_{F}$ 
is located on the strongly decreasing DOS. The obtained $n(E_{F})$ value 
is as large as 80.8 states/Ry/f.u. (21.3 states/Ry for the Fe-site 
against 45.2 states/Ry for the Mn-site in YMnSi), which is below the 
Stoner limit (Table~\ref{table3}). Indeed, a highly accurate spin-polarised 
KKR computations also supported the non-magnetic ground state after a 
very slow convergence. This result is in very good accordance with 
the absence of magnetic moment on the iron site in all $R$FeSi 
series as deduced from $^{57}$Fe M\"{o}ssbauer studies \cite{welterphd}. 

Bearing that YFeSi is characterized by $E_{F}$ located in the vicinity 
of large Fe $d$-DOS in mind, one suggests that this compound may 
be close to a magnetic instability. 
Thus, our theoretical results for YFeSi can not be immediately 
extrapolated to the other $R$Fe$X$ compounds and electronic structure 
of each $R$Fe$X$ compound has to be analysed individually. 
It is interesting to remind that non-magnetic properties of the 
structurally related ThCr$_{2}$Si$_{2}$-type LaFe$_{2}$Ge$_{2}$ compound 
have been previously supported by the LMTO calculations \cite{ishida27}.

Similarly, high DOS at $E_{F}$ has been computed in YCrSi (using extrapolated 
crystallographic parameters $a \approx 3.80$ \AA, $c \approx 7.10$ \AA, 
$z_{Y} \approx$ 0.65, $z_{Si} \approx$ 0.24), which also indicates that 
this compound is close to magnetic instability. 
Nevertheless, up to now, all attempts to stabilize $R$CrSi 
compounds have always failed.

Noteworthy, the ThCr$_{2}$Si$_{2}$-type YCr$_{2}$Si$_{2}$ compound was shown 
to exhibit a magnetic ordering from the neutron diffraction 
measurements \cite{ijjaali28}.

\section{Neutron diffraction results in $R$Mn$X$ silicides and germanides ($R$ = Mg-Ba)}

The low temperature magnetic structures of the investigated $R$Mn$X$ 
compounds consist in antiferromagnetic (001) Mn planes coupled either 
ferromagnetically (MgMnGe, CaMnSi and CaMnGe \cite{welter4,welter7,dascou10}, 
AFI structure) or antiferromagnetically (SrMnGe and BaMnGe \cite{dascou10}, 
AFII structure) as shown in Fig.~\ref{fig4}. We can notice that there exist 
some discrepancies between various experimental studies concerning both 
the reported magnetic structure and the refined value of Mn magnetic moment 
(Table~\ref{table4}). 

Different orientations for the direction of the Mn magnetic moments 
have been reported for MgMnGe: either within the (001) plane \cite{welter7} 
or along the $c$-axis \cite{dascou10}. The two magnetic models lead 
to different moment values at T= 2 K namely $3.3~\mu_{B}$ and $2.9~\mu_{B}$, 
respectively. Similarly, 
in the case of CaMnGe the magnetic moments are along [001] according 
to \cite{dascou10} ($\mu_{Mn} = 3.80~\mu_{B}$), whereas they are tilted 
of about $34^{o}$ 
from the $c$-axis according to \cite{welter4} ($\mu_{Mn} = 3.34~\mu_{B}$). 
Some discrepancy also exists concerning the Mn magnetic moment value 
in CaMnSi. Although the Mn moments have been found to be aligned 
along the $c$-axis in both studies, their amplitudes are 
either $2.84~\mu_{B}$ \cite{dascou10} or $3.27~\mu_{B}$ \cite{welter4}. 

The largest discrepancy between experimental works occurs in the case 
of BaMnGe. The Mn magnetic moment value has been refined to only $1.85~\mu_{B}$ by Dascoulidou et al. \cite{dascou10} (a value strongly underestimated if 
comparing with other $R$Mn$X$ compounds) next corrected to $3.70~\mu_{B}$ 
by Welter et al. in a further study \cite{welter7}. According to these authors, 
the discrepancy simply results from an error in the normalization 
of the magnetic and nuclear intensities during the use of the Fullprof 
refinement software.

No neutron diffraction experiment has been carried out for the YMnSi 
compound. However, according to bulk magnetic measurements \cite{kido3}, a 
spontaneous magnetization is observed above 190 K ($M_{s} = 1.14 \mu_{B}$) 
whereas an antiferromagnet is stabilized at 2 K. 
Later, a neutron study of the La$_{1-x}$Y$_{x}$MnSi solid solution 
\cite{ijjaali9} 
has concluded that a collinear ferromagnetic structure is likely 
for YMnSi above 200 K whereas antiferromagnetically coupled 
ferromagnetic (001) Mn planes characterize the 2K ordering. 
The La$_{0.2}$Y$_{0.8}$MnSi alloy was determined to be a pure ferromagnet 
with a Mn magnetic moment of $2.1(2)~\mu_{B}$ at 2K \cite{ijjaali9}.

Remark : {\it It is worth noting that, in all neutron studies, the 
amplitude of the Mn moment are given with relatively large standard 
deviations (Table~\ref{table4}). As shown in the published neutron 
diffraction patterns and according to the magnetic structures
\cite{welter7,dascou10}, this is due to the lack of magnetic 
contributions which, furthermore, mostly occur under strong nuclear peaks}. 

\subsection{Spin-polarised DOS}

\subsubsection{$R$Mn$X$ ($R$=Mg-Ba)}

Based on the aforementioned experimental results 
\cite{welter4,welter7,ijjaali9,dascou10} the ground state 
electronic structure calculations were performed 
considering, first, a simple ferromagnetic state and next, 
accounting for the real magnetic ordering of Mn sublattices. 
We present DOS in ferromagnetic (Fig.~\ref{fig5a}) and antiferromagnetic 
states (Fig.~\ref{fig5b}), i.e. in the AFI structure for MgMnGe, CaMnGe, 
CaMnSi and in the AFII structure for SrMnGe and BaMnGe. 
Noteworthy, in the latter compounds the $c$ lattice constant was doubled 
to allow for calculations of the AFII structure (Fig.~\ref{fig4}).

First, let us compare KKR total energy computed in non-magnetic and 
different types of magnetic structure. We can roughly conclude (without 
presenting detailed analysis of all contributions to the total energy 
of the system) that our calculations always resulted in a preference 
of a magnetic state in all investigated $R$Mn$X$ compounds. Moreover, 
one observes much lower DOS at $E_{F}$ in the AF state than in the F one 
(Table~\ref{table5}). Undoubtedly, a magnetic structure preference 
should be addressed with a more detailed total energy analysis 
({\it allowing for relaxing of atom positions and lattice parameters}), 
which is however a difficult task for the CeFeSi-type structure 
({\it four adjustable parameters}). Nevertheless, comparing only 
the $n(E_{F})$ values calculated in the three magnetic structure 
types (Table~\ref{table5}), one can verify that the lowest $n(E_{F})$ 
values are computed for the experimentally observed Mn magnetic moment 
arrangement ({\it except for BaMnGe}). 

The density of states at the Fermi level almost vanishes for AFII-SrMnGe 
(Table~\ref{table5}), which predicts electron transport properties close 
to the metal-semiconductor limit. 
The dispersion curves $E({\bf k})$ (Fig.~\ref{fig6}) show an energy 
gap at $E_{F}$ along many directions in the Brillouin 
zone. The main contributions to the finite $n(E_{F})$ come from the valence 
and conduction band overlap near the centre of the Brillouin zone and 
also along the $\Gamma - Z$ direction. 

It seems interesting to remind that electronic structure calculations 
have already suggested similar electron transport behaviour in the 
ThCr$_{2}$Si$_{2}$-type BaMn$_{2}$Ge$_{2}$ compound \cite{tobola29}. 
In order to verify these theoretical predictions, electrical resistivity 
measurements are in progress.

Inspecting site-decomposed and $l$-decomposed contributions to the total 
DOS (Fig.~\ref{fig5b}), one observes that the Mn $d$-states are strongly 
polarised in all cases. The calculated value of the Mn magnetic moment 
varies from $2.81~\mu_{B}$ (MgMnGe) to $3.40~\mu_{B}$ (BaMnGe). 
It can be also noticed that AF coupling in the (001) Mn plane 
substantially increases Mn magnetic moment with respect to the 
F arrangement (Table~\ref{table5}) and this increase of $\mu_{Mn}$ 
is accompanied by lowering of density of states in the vicinity 
of $E_{F}$ (see, Fig.~\ref{fig5a} and Fig.~\ref{fig5b}). 

According to Fig.~\ref{fig1} and Fig.~\ref{fig4} the AFI and AFII Mn 
magnetic arrangements yield zero molecular fields at $R$ and $X$ sites. 
Indeed, no polarization is observed on the $R$ and $X$ site-projected 
DOS from our calculations.

Keeping in mind some limitations of the $muffin-tin$ 
potential in description of electronic structure and magnetism 
of the pseudo-layered $R$Mn$X$ compounds, the agreement between 
theoretical and experimental values is satisfying 
(Table~\ref{table4}). 
Noteworthy, the KKR value obtained for MgMnGe agrees with the 
easy-axis model of Dascouliscou et al. ($\mu_{Mn} = 2.9(2) \mu_{B}$ 
\cite{dascou10}), whereas for CaMnGe the theoretical value 
is close to the experimental one obtained for the easy plane 
arrangement \cite{welter4} (Table~\ref{table4}).

\subsubsection{La$_{1-x}$Y$_{x}$MnSi and 'puzzling' YMnSi}

In order to better understand the magnetic properties of YMnSi, 
KKR-CPA calculations were performed in the ferromagnetic 
La$_{0.2}$Y$_{0.8}$MnSi compound \cite{ijjaali9}. 
The computed Mn magnetic moment 
(2.01 $\mu_{B}$) is in good agreement with the low temperature 
neutron diffraction study (2.1(2) $\mu_{B}$). Due to their ferromagnetic 
environment, the Y/La and Si atoms also exhibit weak DOS polarization and 
experience non-zero molecular field ($\mu_{Y}$= -0.06 $\mu_{B}$ and 
$\mu_{Si}$= -0.08 $\mu_{B}$). 

Inspecting KKR-CPA DOS in Fig.~\ref{fig7} one observes $E_{F}$ 
on the verge of the large DOS peak for the spin-down electrons, 
whereas in the minimum for the spin-up DOS. Hence, if the Y concentration 
tends to $x=1$ in La$_{1-x}$Y$_{x}$MnSi, a slight decrease of 
spin-polarization of the Mn $d$-states can be expected due to the 
unit cell contraction. 
Consequently, the Fermi level would fall into a large spin-down DOS, 
substantially increasing the total energy. Such an electronic 
structure behaviour seems to be responsible for the F-AF transition 
occurring between $x=0.8$ and $x=1.0$ in the La$_{1-x}$Y$_{x}$MnSi 
solid solution \cite{ijjaali9} and detected in YMnSi \cite{kido3}. 

Since the low temperature magnetic structure of the YMnSi compound 
is unknown, the KKR calculations were carried out considering three kinds of 
magnetic orderings: F, AFI and a hypothetical AFIII (e.g. Mn (001) 
ferromagnetic planes antiferromagnetically coupled).

Using the $n(E_{F})$ value as criterion for stability, 
the AFIII magnetic ordering seems to be more favourable for YMnSi 
that the Mn (001) antiferromagnetic planes characterizing the other 
$R$Mn$X$ compounds (Table~\ref{table5}). 

In Fig.~\ref{fig8} we observe that the manganese $d$-states show 
significant spin-polarization, yielding a magnetic moment of 
$1.99~\mu_{B}$, $2.29~\mu_{B}$  and $1.85~\mu_{B}$  in F, AFI and 
AFIII states, respectively. So, accounting only for the Mn-Mn distance 
(Table~\ref{table3} and Table~\ref{table5}), the Mn magnetic moment value 
in the YMnSi compound is much smaller (even in AFI state) that those 
calculated for the other $R$Mn$X$ compounds. Thus, as discussed later, 
the Mn-Si distance (2.54~\AA) probably controls 
the Mn magnetic moment magnitude in this compound. 

\subsection{Mn magnetic moment variation in $R$Mn$X$}

The evolution of $\mu_{Mn}$ versus the in-plane interatomic Mn-Mn 
distance ($d_{Mn-Mn}$) is plotted in Fig.~\ref{fig9}. It is clearly 
seen that the calculated moment increases with d$_{Mn-Mn}$, in fair accordance 
with the experimental data. This evolution can be related to the 
increase of the density of states at the Fermi level and to the 
reduction of the Mn 3$d$ bandwidth when d$_{Mn-Mn}$ increases 
\cite{weijs24,johansson25}. Except for YMnSi, the correlation 
between $\mu_{Mn}$ and $d_{Mn-Mn}$ is really good (Fig.~\ref{fig9}). 
The Mn magnetic moment computed in the AFI magnetic structure of 
YMnSi slightly better fits the values obtained for $R$Mn$X$, 
when plotting $\mu_{Mn}$ versus $d_{Mn-X}$ (Fig.~\ref{fig10}). 

In order to estimate the influence of the Mn-$X$ interatomic distance 
on the Mn magnetic moment amplitude in the $R$Mn$X$ compounds, 
the KKR calculations were performed modifying the $z_{X}$ parameter in
MgMnGe (both in F and AFI state) and YMnSi (F). Such modeling 
does not affect the Mn-Mn distance, which only depends on lattice 
constant $a$. The obtained evolution of $\mu_{Mn}$ versus 
$d_{Mn-X}$ is plotted in Fig.~\ref{fig11}. We immediately note that 
$\mu_{Mn}$ strongly varies with $d_{Mn-X}$, from $1.5~\mu_{B}$  
($d_{Mn-Si} = 2.41$ \AA) to about $2.5~\mu_{B}$ ($d_{Mn-Si} = 2.7$ \AA) in 
YMnSi and from $2.5~\mu_{B}$ ($d_{Mn-Ge} = 2.48$ \AA) to about 
$3~\mu_{B}$ ($d_{Mn-Ge} = 2.65$ \AA) in MnMgGe. 

These results apparently suggest that manganese-metalloid distances 
should have an essential effect on the magnetic properties of $R$Mn$X$. 
However, the interplay with the Mn-Mn distance, seen in Fig.~\ref{fig9}, 
should be also underlined. 

Moreover, it is important to mention that integrating the 
spin-polarised $d$-Mn DOS in the $R$Mn$X$ compounds give similar 
number of electrons occupying $d$ orbitals (about 5.6 electrons 
per Mn), whatever the nature of $R$, the Mn-Mn and Mn-$X$ interatomic 
distances and the magnetic structure type. Let us recall that 
previous experimental works suggested that the Mn magnetic moment 
variation within the $R$Mn$X$ family of compounds is rather related 
to subsequent filling of the Mn 3$d$ band (mostly due to the $X$-Mn 
charge transfer) when $d_{Mn-X}$ decreases. In view of KKR results 
the $\mu_{Mn}$ enhancement with $d_{Mn-X}$ is due to the 
variable distribution of the Mn 3$d$-DOS between two spin 
channels (Fig.~\ref{fig12}). 

The analysis of the Mn magnetic moment dependence on Mn-$X$ interatomic 
distances may be decomposed by considering two intimately related 
effects, namely the covalent magnetism effect 
(introduced by Williams et al. \cite{williams34,williams35,kubler33} 
as an alternative to the Stoner model) and the modification of the 
intra-atomic Coulomb repulsion with the $p-d$ hybridisation. 

Recently, based on electronic structure calculations, 
the observed variation of $\mu_{Fe}$ with the $R$ element 
valence in the $R$Fe$_{6}$Ge$_{6}$ compounds was interpreted 
in terms of spin-dependent covalent interactions \cite{mazet36}.

In all investigated $R$Mn$X$ compounds, whatever the $R$ element, 
the spin-up and spin-down Mn 3$d$ sub-bands are not the same and the 
two sub-bands do not shift in rigid way when going from the non-magnetic 
to ferromagnetic state (as expected from the Stoner model). 
The spin-polarised DOS clearly exhibit a spin-dependent hybridisation 
of the Mn 3$d$ states with other electronic states. As already mentioned 
in Sect. 4, in the case of (La-Y)MnSi, the spin-dependent Mn 3$d$ -Y 4$d$ 
hybridisation yields a negative induced magnetic moment on the Y (La) site. 
Moreover, the $X$ $p$ - Mn 3$d$ hybridisation is also spin-dependent, 
and a negative value is calculated for the Si magnetic moment in the 
ferromagnetic La$_{0.2}$Y$_{0.8}$MnSi compound (also observed on $X$-site 
in the F spin-polarised computations carried out in the $R$Mn$X$ series). 
Such result also reflects the mostly spin-down character of the $X$ 3$p$ - 
Mn 3$d$ bonding states at the $X$-site. 
Both behaviours may illustrate the covalent magnetism effect 
\cite{williams35,kubler33}.

In order to show the effect of the Si 3$p$ - Mn 3$d$ hybridisation 
on the shape of the spin-polarised Mn 3$d$ DOS, and consequently on the 
Mn magnetic moment, we compared (Fig.~\ref{fig13}) the non-magnetic and 
ferromagnetic KKR results obtained for three different Mn-Si distances 
in the YMnSi compound. For the sake of clarity, only the Si 3$p$ and 
Mn 3$d$ states were displayed in Fig.~\ref{fig13}. The Mn 3$d$ density 
of states strongly increases in the vicinity of $E_{F}$ when $d_{Mn-Si}$ 
increases. At the same time, the Si 3$p$ - Mn 3$d$ states, presumably 
lying in the antibonding energy range  ($0.10 < E < - 0.15$ Ry), 
shift toward $E_{F}$. The separation between bonding and antibonding 
states decreases, thus indicating a reduction of the $p-d$ hybridisation. 
In the ferromagnetic state, the spin-up and spin-down Mn 3$d$ sub-bands 
evolve in different and complex way and one observes the displacement of 
their energy barycentre toward the low and high energies, respectively. 
Thus, the Fermi level progressively moves into the region with a low 
spin-down DOS, tentatively identified as the bonding-antibonding 
states separation due to strong overlaps between the Mn 3$d$ orbitals. 
Consequently, some electrons are transferred between 3$d$ spin-channels 
(from 'down' to 'up') and the Mn magnetic moment substantially increases.

A qualitative explanation of this behaviour can be also proposed 
using the Kanamori's arguments \cite{kanamori37}. The increase of 
the Mn-$X$ distances is responsible for the decrease in the strength 
of the $p-d$ hybridisation, which results in the increase of the 
amplitude of wave functions on the Mn-site. Consequently, the 3$d$-states 
on Mn become more and more localized and one can expect the increase 
of the repulsive intra-atomic Coulomb energy U (the parallel spin 
arrangement should be favoured to decrease the electrostatic interactions). 
In order to compensate the increase of U, there is an increase in the 
spin-up $d$-states occupation, at the expense of the spin-down $d$-states.

\section{Conclusions}

The KKR electronic structure calculations shed light on 
the magnetic properties of the manganese sublattice in 
the ternary CeFeSi-type $R$Mn$X$ compounds. 

From analysis of the spin-dependent electronic structures, we 
conclude that the magnetic behaviours of these compounds 
can be better described in terms of the covalent magnetism than 
with the Stoner model. The increase of the Mn magnetic moment 
with interatomic distances is due to a redistribution of electrons 
between 3$d$ spin-up and spin-down subbands.

The verification of the calculated electronic structure in the entitled 
compounds could be done by e.g. photoemission method. Recently, the results 
of X-ray photoemission spectroscopy (XPS) have been interpreted 
by the TB LMTO method in the similar YbMn$_{2}X_{2}$ compounds 
\cite{szytula2004}. To our knowledge such experiments have been 
not performed yet for $R$Mn$X$ compounds. 

The influence of the interatomic distances, intraplanar Mn-Mn 
and particularly Mn-$X$, on the polarization of the Mn 3$d$ states 
(the amplitude of $\mu_{Mn}$), well supports the experimentally 
observed evolution in the $R$Mn$X$ series. 

The electronic structure calculations in $R$(=Mg-Ba)Mn$X$ silicides 
and germanides yield $\mu_{Mn}$ in rather good agreement 
with the experimental ones (Fig~\ref{fig14}). 

The computations have also revealed that precise crystallographic 
data (especially atomic coordinates controlling $d_{Mn-X}$ distances) 
are of primary importance to allow for phenomenological predictions of 
magnetic behaviours in these compounds from simple steric 
criteria.

Comparing DOS computed in the F, AFI and AFII states, we have found 
that the lowest value of $n(E_{F})$ corresponds to experimentally 
observed magnetic orderings (i.e. AF (001) Mn planes) in most $R$Mn$X$ 
compounds. Using the same $n(E_{F})$ criterion, in YMnSi 
(high pressure compound), the KKR calculations favour the occurrence 
of ferromagnetic (001) Mn planes (antiferromagnetically coupled along 
the $c$-axis), as suggested in previous experimental works on 
La$_{1-x}$Y$_{x}$MnSi solid solution. The KKR-CPA calculations 
of the ferromagnetic La$_{0.2}$Y$_{0.8}$MnSi compound resulted 
in a Mn magnetic moment value in very good agreement with the 
experimental data. 

In another way, unlike the strong magnetic properties of the 
$R$Mn$X$ compounds, the ground state of YFeSi was determined 
to be non-magnetic, which is in excellent agreement with the 
M\"{o}ssbauer data.

Finally, it would be highly desirable to perform electronic 
structure calculations \cite{brooks38,ericsson39} for 
$R$Mn$X$ compounds 
containing $4f$ element due to a wide variety of their magnetic 
behaviours \cite{welter40,welter41}.
 
\section*{Acknowledgments}
We are indebted to P. P\'{e}cheur (LPM, Ecole des Mines, Nancy) for his 
critical reading of the manuscript.

%
%
\begin{table}
\caption{Crystallographic parameters of the CeFeSi-type structure
(space group P4/$nmm$, $a \approx 4$ \AA, $c \approx 7 \AA$)}.
\label{table1}
\begin{tabular}{cccccc}
\hline
\hline
Atom 	& Position & Point symmetry & x & y & z \\
\hline
Ce	& 2(c)	& $4mm$	& ${1\over4}$ & ${1\over4}$ & $\approx 0.66$ \\	
Fe	& 2(a)	& $42m$	& ${3\over4}$ & ${1\over4}$ & 0 \\
Si	& 2(c) 	& $4mm$	& ${1\over4}$ & ${1\over4}$ & $\approx 0.18$ \\
\hline
\hline
\end{tabular}
\end{table}

\begin{table}
\caption{Lattice parameters (in ~\AA) and atomic positions 
of the CeFeSi-type $R$Mn$X$ compounds. Mn-Mn interatomic distance 
(d$_{Mn-Mn}=\sqrt{2} a$) is given in~\AA.} 
\label{table2}
\begin{tabular}{lllllll}
\noalign{\smallskip}\hline
Compound & a (\AA) & c (\AA) & z$_{R}$ & z$_{X}$ & d$_{Mn-Mn}$ & Ref. \\
\noalign{\smallskip}\hline
MgMnGe & 3.949 & 6.535 & 0.655 & 0.252 & 2.79 & \cite{welter7} \\
CaMnGe & 4.227 & 7.201 & 0.661 & 0.220 & 2.99 & \cite{welter4,dascou10,pearson} \\
SrMnGe & 4.381 & 7.482 & 0.667 & 0.206 & 3.10 & \cite{dascou10,welter11,pearson} \\
BaMnGe & 4.507 & 7.893 & 0.668 & 0.190 & 3.19 & \cite{dascou10,welter11,pearson} \\
CaMnSi & 4.172 & 7.121 & 0.669 & 0.213 & 2.95 & \cite{welter4} \\
\noalign{\smallskip}\hline
YMnSi$^{*}$ & 3.970& 7.150 &0.67 &0.22 &2.81& \cite{kido3} \\
\noalign{\smallskip}\hline
\end{tabular}
$^{*}$ {\small extrapolated atomic coordinates}
\end{table}

\begin{table}
\caption{KKR $d$-DOS at the Fermi level for $T$-site  
(in states/Ry) and Stoner products $I n(E_{F})$ for $R$Mn$X$. 
The corresponding interatomic distances, $d_{T-X}=\sqrt{2}a$ 
and $d_{T-X}=\sqrt{{{a^2}\over4}+z_{X}^{2}c^{2}}$, 
are given in \AA.} 
\label{table3}
\begin{tabular}{lcccc}
\hline
Compound &  $d_{T-T}$ &  $d_{T-X}$ & $n_d(E_F)$ & $I n(E_{F})$ \\
\hline
MgMnGe & 2.79 & 2.57 & 43.5 & 1.39 \\
CaMnSi & 2.95 & 2.58 & 58.6 & 1.81 \\
CaMnGe & 2.99 & 2.64 & 64.2 & 1.96 \\
SrMnGe & 3.10 & 2.67 & 72.8 & 2.26 \\
BaMnGe & 3.19 & 2.71 & 78.2 & 2.40 \\
\hline
YCrSi$^{*}$  & 2.76 & 2.55 & 21.5 & 0.72 \\
YMnSi$^{**}$  & 2.81 & 2.54 & 43.6 & 1.45 \\
YFeSi$^{**}$  & 2.75 & 2.32 & 19.6 & 0.68 \\
\hline
\end{tabular}
\\
$^{*}$ {\small hypothetical structure} \\
$^{**}$ {\small extrapolated atomic coordinates}
\end{table}

\begin{table}
\caption{Experimental and calculated Mn magnetic moments 
(in $\mu_{B}$) for antiferromagnetic $R$Mn$X$ compounds.} 
\label{table4}
\begin{tabular}{llcc}
\hline
Compound &  $\mu_{exp}$  & $\mu_{cal}$  & Ref. \\
\hline
MgMnGe 	& 2.9(2)  $^{*)}$ & 2.81 & \cite{dascou10}	\\
	& 3.3(1)  $^{**)}$ & 	& \cite{welter7}	\\
CaMnGe 	& 3.34(3)  tilted & 3.15 & \cite{welter4}	\\
	& 3.8(2)  $^{*)}$ & 	& \cite{dascou10}	\\
SrMnGe  & 3.29(6)  $^{**)}$ & 3.31 & \cite{dascou10}	\\
BaMnGe  & 3.70(6)  $^{**)}$ & 3.40 & \cite{dascou10}	\\
CaMnSi  & 2.84(8)   $^{*)}$ & 3.02 & \cite{dascou10}	\\
	& 3.27(4)   $^{*)}$ &	& \cite{welter7}	\\
\hline
\end{tabular}
\\
$^{*}$ {\small parallel $c$-axis} \\
$^{**}$ {\small perpendicular $c$-axis} \\
\end{table}

\begin{table}
\caption{KKR results, Mn magnetic moment $\mu_{Mn}$ (in $\mu_{B}$) and 
DOS at $E_{F}$ $n(E_{F})$ (in states/Ry/f.u.) in the $R$Mn$X$ compounds 
in ferromagnetic (F) and antiferromagnetic (AFI, AFII) states. 
The values in bold correspond to the results obtained accounting for 
experimentally observed magnetic structure.} 
\label{table5}
\begin{tabular}{llllrrr}
\hline
Compound & \multicolumn{3}{c}{$\mu_{Mn}$} & \multicolumn{3}{c}{$n(E_{F})$} \\
\hline
	& F & AFI & AFII & F & AFI & AFII \\
\hline
MgMnGe	& 2.19 	& {\bf 2.81}	& 2.69		& 70.5	& {\bf 63.4}	& 70.5	\\
CaMnGe	& 2.50 	& {\bf 3.15}	& 3.15		& 129.6	& {\bf 21.7}	& 24.7	\\
SrMnGe	& 3.09 	& 3.33		& {\bf 3.31}	& 105.8	& 104.1	& {\bf 0.1}	\\
BaMnGe	& 3.22 	& 3.43		& {\bf 3.40}	& 115.9	& 3.5	& {\bf 15.3}	\\
CaMnSi	& 2.06 	& {\bf 3.02}	& 2.59		& 109.7	& {\bf 33.1}	& 105.6	\\
\hline
YMnSi	& 1.99 	& 2.29		& 1.85 $^{*}$	 	& 91.9	& 81.2 	& 66.1	\\
\hline
\end{tabular}
$^{*}$ {\small AFIII ordering}
\end{table}

%
%
\newpage
\begin{figure}
\caption{Tridimensional views of the CeFeSi-type and ThCr$_{2}$Si$_{2}$-type 
structures} 
\label{fig1}
\end{figure}

\newpage
\begin{figure}
\caption{KKR non-spin-polarised DOS for the $R$Mn$X$ compounds. 
The upper panel is total DOS. The $s$, $p$ and $d$ contributions are 
plotted by dotted, solid thick and solid thin lines, respectively.}
\label{fig2}
\end{figure}

\begin{figure}
\caption{KKR non-polarised total and $T$-site DOS for Y$T$Si 
($T$= Cr, Mn, Fe).}
\label{fig3}
\end{figure}

\begin{figure}
\caption{Tridimensional views of the magnetic structures of the 
CeFeSi-type $R$Mn$X$ silicides and germanides.}
\label{fig4}
\end{figure}

\begin{figure}
\caption{KKR spin-polarised DOS for the ferromagnetic $R$Mn$X$ 
compounds.}
\label{fig5a}
\end{figure}

\begin{figure}
\caption{KKR spin-polarised DOS for the antiferromagnetic 
$R$Mn$X$ compounds.}
\label{fig5b}
\end{figure}

\begin{figure}
\caption{The electronic dispersion curves E({\bf k}) in the SrMnGe 
antiferromagnet. For a sake of clarity only bands in the vicinity of 
the Fermi level were plotted.} 
\label{fig6}
\end{figure}

\begin{figure}
\caption{KKR-CPA spin-polarised DOS for La$_{0.2}$Y$_{0.8}$MnSi. 
Mn-DOS is plotted by dotted line.} 
\label{fig7}
\end{figure}

\begin{figure}
\caption{KKR spin-polarised DOS for YMnSi in F, AFI and AFIII states. 
Mn-DOS is plotted by a thick line.} 
\label{fig8}
\end{figure}

\begin{figure}
\caption{The calculated Mn magnetic moment versus the Mn-Mn distance 
in $R$Mn$X$. A line is added as a guide to the eye.}
\label{fig9}
\end{figure}

\begin{figure}
\caption{The calculated Mn magnetic moment versus the Mn-$X$ 
distance in $R$Mn$X$. A line is added as a guide to the eye.}
\label{fig10}
\end{figure}

\begin{figure}
\caption{Modeling of the Mn magnetic moment variations with the Mn-$X$ 
distance in YMnSi (in F state) and MgMnGe (in AFI state). 
Only $z_{X}$ was modified in the KKR simulations.}
\label{fig11}
\end{figure}

\begin{figure}
\caption{Number of electrons filling spin-up and spin-down Mn $d$-DOS 
versus the Mn-Mn distances in the $R$Mn$X$ compounds. Lines are added 
as a guide to the eye.}
\label{fig12}
\end{figure}

\begin{figure}
\caption{KKR non-spin-polarised (left panel) and spin-polarised 
(right panel) DOS in the CeFeSi-type YMnSi simulated at three different 
Mn-Si interatomic distances ($d_{Mn-Si}$, given in \AA). 
For a sake of clarity only Mn $d$-states (thin line) and Si 
$p$-states (thick line) are plotted.}
\label{fig13}
\end{figure}

\begin{figure}
\caption{Comparison of experimental (with error bars) and theoretical 
values of the Mn magnetic moment in $R$Mn$X$ compounds (see text).}  
\label{fig14}
\end{figure}

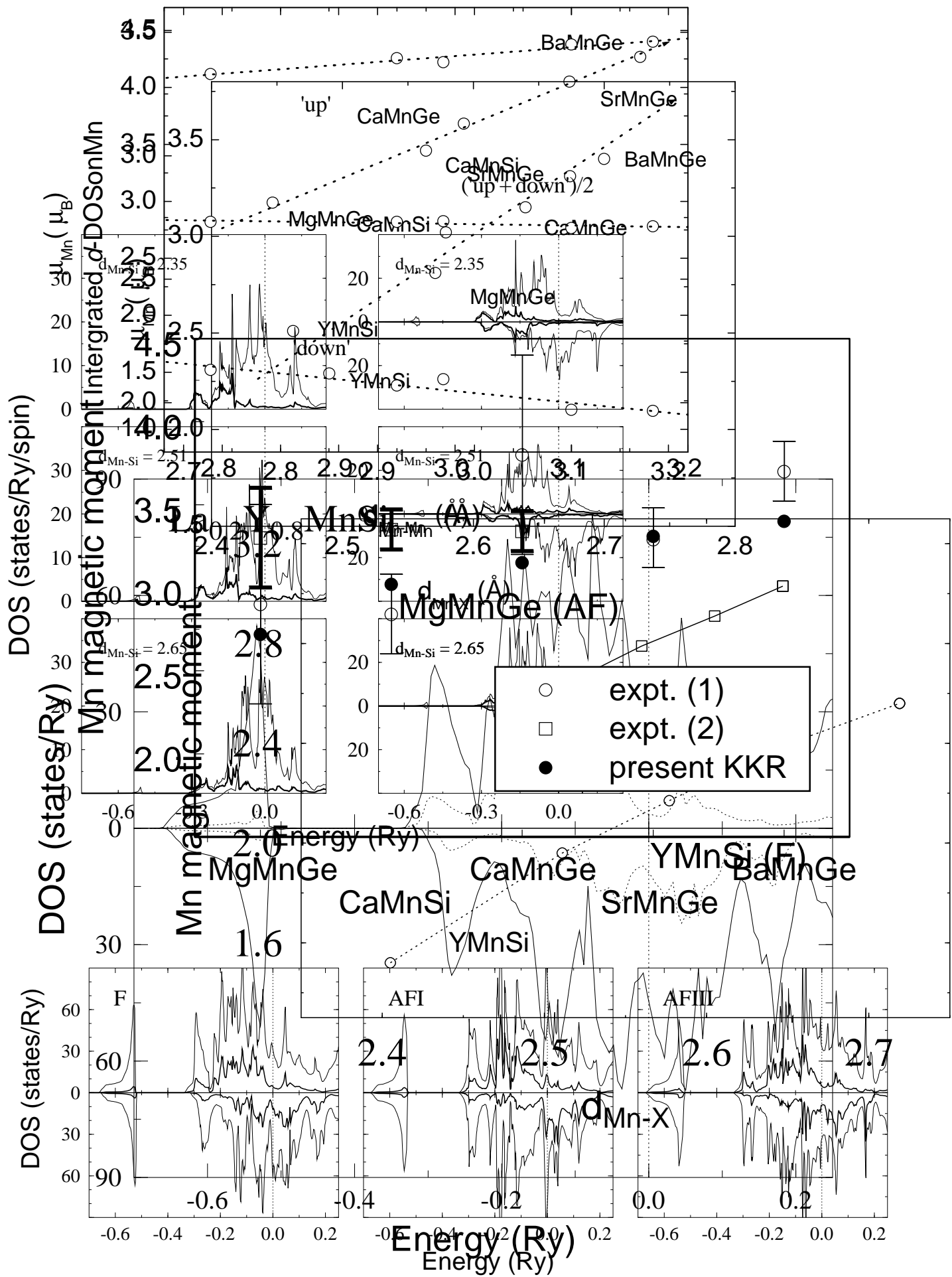
\end{document}